# Diagnosing epilepsy using entropy measures and embedding parameters of EEG signals

Fatemeh Valipour [a], Zahra Valipour [b], Mani Garousi [c], Ali Khadem [a,*]

[a] Department of Biomedical Engineering, Faculty of Electrical Engineering, K. N. Toosi University of Technology, Tehran, Iran

[b] Department of Biomedical Engineering, Faculty of Advanced Technologies in Medicine, Isfahan University of Medical Sciences, Isfahan, Iran

[c] Department of Biomedical Engineering, Faculty of Electrical and Computer Engineering, Tarbiat Modares University, Tehran, Iran

[*] Corresponding author address: Ali Khadem, Ph.D.Assistant Professor, Department of Biomedical Engineering, Faculty of Electrical Engineering (EE), K. N. Toosi University of Technology (KNTU), Tehran 1631714191, Iran, P.O. Box 16315-1355, Tel.: +98-21-84062450 (204)

E-mail: alikhadem@kntu.ac.ir

## Abstract

Epilepsy is a neurological disorder that affects normal neural activity. These electrical activities can be recorded as signals containing information about the brain known as Electroencephalography (EEG) signals. Analysis of the EEG signals by individuals for epilepsy diagnosis is subjective and time-consuming. So, an automatic classification system with high detection accuracy is required to overcome possible errors. In this study, the discrete wavelet transform has been applied to EEG signals. Then, entropy measures and embedding parameters have been extracted. These features have been investigated individually to find the most discriminating ones. The significance level of each feature was evaluated by statistical analysis. Consequently, LDA and SVM algorithms have been employed to categorize the EEG signals. The results have indicated that the features of Embedding parameters, PermutationEntropy, FuzzyEntropy, SampleEntropy, NormEntropy, SureEntropy, LogEntropy, and ThresholdEntropy have the potential to discriminate epileptic patients from healthy subjects significantly. Also, SVM classifier has achieved the highest classification accuracy. In this study, we could find effective embedding-based and entropy-based features as appropriate single measures for identifying abnormal activities that can efficiently discriminate the EEG signals of epileptics from healthy individuals. According to the results, they can be used for automatic classification of epileptic EEG signals that are difficult to examine visually.

## Introduction

Epilepsy is one of the common disorders of the central nervous system caused by brain electrophysiological system malfunctions. Many people suffer from this disease all over the world. Recurrent seizures are one of the symptoms of epilepsy that occur in unpredictable times, usually without warning. Seizures can lead to attention loss, whole-body convulsion or even death. Therefore, it is necessary to be quantitatively evaluated [1]. Electroencephalography (EEG) is the most popular noninvasive modality for detecting and monitoring neurological disorders. It measures the electrical activity of the brain by using electrodes. The electrodes are evenly arranged on the scalp. Every EEG channel records the sum of the postsynaptic potentials of a large number of neurons. EEG has a high temporal resolution. It measures the electrical activity of the brain directly, and is a useful clinical tool, especially in the field of epileptology. One can analyze and examine the EEG signal visually. However, that analysis could be subjective, time-consuming and requires skilled personal interpretations which may be prone to errors and also small details of EEG signals including important information may be ignored [2, 3]. Therefore, to overcome these constraints, an automatic system for detecting the epileptic EEG signals is required.

Major studies about epileptic seizure detection are summarized in Table 1.

Table 1. Summary of previous studies for diagnosing epilepsy by EEG:

| Author | Feature | Classifier | EEG dataset | Accuracy (%) | Specificity (%) | Sensitivity (%) |
|---|---|---|---|---|---|---|
| Subasi et al [4] | SD and average value | LR<br>MLPNN | [4] | -------<br>------- | 90.3<br>91.4 | 89.2<br>92.8 |
| Subasi et al [6] | Statistical features with<br>ICA,<br>PCA,<br>LDA | SVM | University of Bonn [5] | 99.50<br>98.75<br>100 | 99<br>98.5<br>100 | 100<br>99<br>100 |
| Jia et al [7] | Statistical features in the CEEMD domain | Random forest | University of Bonn [5] | 98 | 99 | 100 |
| Donos et al [9] | Time-domain and power band features | Random forest | EMU [8] | ------- | ------- | 86.27 |
| Faust et al [10] | Frequency domain parameters, Burg's method | SVM | University of Bonn [5] | 93.3 | 98.33 | 96.67 |
| Polat et al [11] | Fourier transform-based features | Decision tree | University of Bonn [5] | 98.72 | 99.31 | 99.40 |
| Subasi et al [12] | AR parameter estimation and maximum likelihood estimation | Wavelet neural networks and back propagation | [12] | 93 | 92.4 | 93.6 |
| Ubeyli et al [13] | AR method | SVM | University of Bonn [5] | 99.56 | 99.63 | 99.50 |
| Chandaka et al [14] | Cross-correlation | SVM | University of Bonn [5] | 95.96 | 100 | 92 |
| Orhan et al [15] | Wavelet-based features | K-means clustering and MLP | University of Bonn [5] | 96.67 | 97.98 | 94.12 |
| Shengkuh et al [16] | Wavelet variances | KNN | University of Bonn [5] | 100 | ------- | ------- |
| Guo et al [17] | Relative Wavelet Energy | ANN | University of Bonn [5] | 95.2 | 92.12 | 98.17 |
| Wang et al [18] | Wavelet packet entropy | KNN | University of Bonn [5] | 99.44 | ------- | ------- |
| Redilico et al [19] | ReniEn<br>TallisEn | LR | University of Bonn [5] | 95<br>94.5 | 94<br>94 | 97<br>97 |
| Kannathal et al [20] | SpectralEn, EmbeddingEn, Kalmogorov—SinaiEn, ApproximateEn | ANFIS | University of Bonn [5] | 92.2 | ------- | ------- |
| Gupta et al [22] | Cross corrEn, log energy En, SureEn | Least square-SVM (RBF kernel)<br>KNN(Euclidean distance) | Bern Barcelona database [21] | 94.41<br>93.12 | 95.57<br>95.15 | 93.25<br>91.09 |
| Gandhi et al[23] | DWT+(Spectral entropy ,Energy) | PNN | University of Bonn [5] | 100 | ------- | ------- |
| Guo et al [24] | Multi wavelet transform+ ApproximateEn | MLPNN | University of Bonn [5] | 98.2 | 95.50 | 99.00 |
| Das et al [25] | EMD-DWT method, log-energy entropy | KNN (Cityblock distance) | University of Bonn [5] | 89.4 | 88.1 | 90.7 |
| Joshi et al [26] | Fractional linear prediction | SVM (RBF kernel) | University of Bonn [5] | 96 | 95 | 95.33 |
| Polychronaki et al [27] | Fractal dimension | KNN | Epilepsy Telemetry Unit, Department of Neurosurgery, University of Athens, 'Evangelismos' Hospital | ------- | ------- | 100 |
| Chua et al [28] | Higher order statistics based features | GMM | University of Bonn [5] | 93.1 | 92 | 97.67 |
| Gruszczyńska et al [30] | RQA measures | SVM | Medical University of Bialystok [29] | 86.8 | ------- | ------- |
| Acharya et al [31] | RQA measures | GMM<br>KNN | University of Bonn [5] | 92.6<br>95.2 | 92.2<br>98.9 | 97.2<br>98.3 |
| Guler et al [32] | Lyapunov exponent | RNN | University of Bonn | 96.79 | ------- | ------- |
| Acharya et al [33] | HOS+Higuchi FD+Hurst EXPONENT+AppEn+SampEn | DT,<br>PNN,<br>KNN,<br>Fuzzy,<br>GMM,<br>SVM | University of Bonn [5] | 97.3<br>98.1<br>98.1<br>100<br>99<br>99 | 94<br>96<br>96<br>100<br>98<br>98 | 99<br>99<br>99<br>100<br>100<br>99.5 |
| Acharya et al [34] | | Deep convolutional neural network | University of Bonn [5] | 88.67 | 90 | 95 |

SD, Standard Deviation; LR, Logistic regression; MLPNN, Multilayer Perception Neural Networks; AR, Autoregressive; ANN, Artificial Neural Network; k-NN, K-Nearest Neighbor; LS, Least Squares; SVM, Support Vector Machine; GLM, Generalized Linear Model; HOS, Higher Order Spectra; PNN, Probabilistic neural network; ASE, Average Sample Entropy; AVIF, Average Variance of Instantaneous Frequencies; DWT, Discrete Wavelet Transform; AppEn, Approximate Entropy; RQA, Recurrence Quantification Analysis; RNN, Recurrent Neural Networks; EMD, Empirical Mode Decomposition; GMM, Gaussian mixture model; ApEn, Approximate Entropy;

Separation of epileptic patients from the healthy group is a pattern recognition problem which its most important step is feature extraction. The quality of the selected features is an effective factor in diagnostic performance.

Based on Table 1 the features used for epilepsy diagnosis can be divided into different groups based on different perspectives:

- Time domain, frequency domain, and time-frequency domain features
- Linear and nonlinear features
- Activity, connectivity, and complexity features

Most biological signals and especially EEG are essentially nonlinear and non-periodic. Therefore, in many cases, linear analysis (e.g. FFT) is not able to differentiate EEG signals of disordered subjects from healthy ones. Consequently, nonlinear methods should be considered [35].

State-space reconstruction is the dynamic analysis for efficient estimation of embedding measures. The time series dynamic is presented by the state-space reconstruction [30]. The phase-space reconstruction methods have been successfully used in analyzing EEG signals [36-38], however, to the best of our knowledge and based on Table 1, they haven't been used for differentiating EEG signals between epileptic and healthy subjects. Therefore, it can be worthy to investigate the feasibility of the reconstructed phase space for each normal and epileptic EEG signal and evaluate the discrimination provided by embedding parameters.

In addition, although different entropy measures have been used in the literature for epileptic EEG detection (See Table 1) to the best of our knowledge no study compared the power of various kinds of entropy measures and different classifiers to classify epileptic EEG signals on the same EEG dataset.

In the present study, we will investigate the accuracy of embedding delay, embedding dimension, and different entropy measures by using LDA and SVM classifiers for epileptic EEG signals from the normal ones.

This paper is organized as follows: In the next section, first, the epileptic EEG data used in this study will be described. Afterwards, the features and classifiers that are investigated in this paper will be introduced. Subsequently, our approach will be proposed. Then, the results will be presented and discussed, respectively. Finally in section V, the paper will be concluded and some future works will be proposed.

## Methods

### EEG data

The EEG signals were collected by Andrzejak et al. [5] at the University of Bonn, Germany. All EEG signals were recorded with the same 128-channel amplifier system. EEG data includes five sets of EEG segments obtained from five healthy subjects with eyes-closed (Set Z) and eyes-open (Set O) and five patients with the stages of interictal (Set N), ictal (Set F) and seizure (Set S), each of which contained 100 single-channel EEG segments. The sampling rate and duration time were 173.6 and 23.6 seconds, respectively. In this study, two sets of the dataset were selected, including the healthy group with open eyes (Set O) and epileptic seizure patients group (Set S).

### Definition of selected features

#### State-space reconstruction

The analysis of nonlinear dynamic systems can be described by a point in the phase space at each instant (e.g. $R$). Phase space is a mathematical space constructed by the dynamic variables of the system. If there are $n$ variables in the dynamic space, a point in the Euclidean space represents the $R^n$ state at any time. By changing the values of these variables, this point moves through the phase space and forms the system attractor.

The state-space reconstruction is calculated by using the Takens embedding method. In this method, we considered time series of $x(n)$; $n=1,2,\dots,N$, and then the state space vectors were constructed by delayed time series samples as follows :

$$\overrightarrow{X_t} = (x(t), x(t+\tau), x(t+2\tau), \dots, x(t+(m-1)\tau)) \quad t = 1,\dots, N-(m-1)\tau \quad (1)$$

Where $X_t$ is reconstructed state space vector, $m$ is embedding dimension and $\tau$ shows the time delay.

The important point in state-space reconstruction is determining the optimal time delay and embedding dimension. For the optimum value of time delay, each independent axis in the m-dimensional space contains the signal information in each direction with the least correlation with the other directions, and these trajectories do not intercept each other [39]. There are several ways to select $\tau$ and $m$. In this study, we use the first minimum of average mutual information and false nearest neighbor method to select $\tau$ and $m$, respectively [40].

Average Mutual Information (AMI)

Fraser and Swinney believed that the first local minimum of average mutual information function is the optimum value of time delay. The average mutual information represents the amount of predictable information of a pointin a time series.

We obtained the data between the two time series for different values of $\tau$ as follows:

The point where the first minimum of $I(\tau)$ occurs is chosen to be as optimal value of $\tau$ [40].

$$I(\tau) = \sum_{n-1}^{N-\tau} P(x(n), x(n+\tau)) \times \log\left(\frac{P(x(n), x(n+\tau))}{P(x(n)) \times P(x(n+\tau))}\right) \quad (2)$$

where $P(.)$ denotes the probability density function.

False Nearest Neighbor (FNN)

In this method, the distance between two points in phase space is examined when a consecutive increment in spatial dimension occurs. If the distance of these two points in a space with dimension $D$ is significantly different from space with the dimension $D+1$, these two points are considered as false neighbors. The distance between the $X(t)$ and its $r^{th}$ nearest neighbor ($x(t_r)$) in the dimensions $D$ and $D+1$ are estimated as follows [40]:

$$R_D^2(t,r) = \| X(t) - X(t_r) \|^2 = \sum_{k=0}^{D-1} [x_{t-k\tau} - x_{t_r-D\tau}]^2 \quad (3)$$

$$R_{D+1}^2(t,r) = R_D^2(t,r) + [x_{t-D\tau} - x_{t_r-D\tau}]^2 \quad (4)$$

If the following conditional equation is satisfied, $x(t_r)$ will be considered as False Nearest Neighbor of $x(t)$, where $R_{tot}$ is a threshold value.

$$\left[\frac{R_{D+1}^2(t,r) - R_D^2(t,r)}{R_D^2(t,r)}\right]^{\frac{1}{2}} > R_{tot} \quad (5)$$

Entropy

Generally, entropy is a nonlinear measure that characterizes the level of signal complexity. The reduction of entropy makes the time series more regular, indicating an increase in the information rate contained in the signal [41]. Approximate entropy, Sample entropy and Fuzzy entropy are based on average logarithm of signal conditional probability such that when the two sequences of the time series are similar at $m$ points, they remain similar at $m + 1$ points.

In the following, we briefly introduce 10 types of entropies:

Approximate entropy (AppEn)

In order to find a specific pattern of the time series, we establish a relationship between the probabilities that is called approximated entropy (AppEn), in which the degree of similarity between the different parts of the signal is measured by a similarity criterion (within a tolerance *r*), and thus the complexity level is quantified [42]. AppEn is calculated as follows:

$$C_i^m(r) = n_i^m(r) / (N - m + 1) \tag{6}$$

$$\varphi^m(n) = (N - m + 1)^{-1} \sum_{i=1}^{N-m+1} \ln C_i^m(r) \tag{7}$$

$$ApEn(m, r) = \varphi^m(r) - \varphi^{m+1}(r) \tag{8}$$

Where $n_i^m(r)$ is the number of m-dimensional similar patterns at a distance less than *r*.

Sample Entropy (SampEn)

SampEn is a modified version of AppEn which in contrast with AppEn is less dependent on the length of the signal, does not count self matches and is unbiased. So it is a more reliable measure [43].

The following equation determines SampEn:

$$SampEn(m, r, N) = -\ln(C^{m+1}(r) / C^m(r)) \tag{9}$$

Where $C^m(r)$, is the number of m-point sequences which remained similar in the distances less than *r*.

Fuzzy Entropy (FuzzyEn)

FuzzyEn is a relatively new method that measures the fuzziness and uncertainty of the time series when the similarity degree of two vectors is fuzzily defined. It is independent of data length and is calculated as follows [44]:

$$\varphi^m(n, r) = [1 / (N - m)] \sum_{i=1}^{N-m} ([\frac{1}{N - m + 1}] \sum_{j=1, j \neq i}^{N-m} D_{ij}^m \tag{10}$$

$$FEn(m, n, r, N) = \ln(\varphi^m(n, r) - \varphi^{m+1}(n, r)) \tag{11}$$

Where $D_{ij}^m$ identifies the similarity degree of two vectors with distance less than *r*, dimension *m*, and gradient boundary *n*.

In this research values of parameters *n*, *r* and *m* are chosen to be 2, 0.15 of the standard deviation of time series and 2, respectively.

Shannon entropy (ShanEn)

This entropy is a measure of a set of relational parameters that changes linearly with the logarithm of the number of probabilities. This can be expressed as:

$$ShanEn(x) = -\sum_k P(x_k) \times \log_2(P(x_k)) \qquad (12)$$
$$0 \langle M \langle N - 1$$

Where $P(x_k)$ is the probability of $x_k$ and $M$ denotes the number of levels of the discrete-valued random variable X.

Spectral entropy (SEn)

Spectral entropy is a normalized form of Shannon entropy that uses the amplitude components of the time series power spectrum to evaluate entropy. This entropy can quantify the spectral complexity of the signal. It is calculated as follows [3]:

$$SEn = -\sum_f p_f \times \log(p_f) \qquad (13)$$

Permutation Entropy (PermEn)

Permutation Entropy estimates the complexity by identifying couplings between time series. This measure quantifies the presence or absence of permutation patterns of the variables in the signal time series.

Suppose the time series $x$ with an embedding dimension of $m$ and time delay of $\tau$ is made as follows [45]:

$$X(N-(m-1)\tau) = \{x(N-(m-1)\tau), x(N-(m-2)\tau), ..., x(N)\} \qquad (14)$$

PermEn is given by:

$$\text{PermEn} = -\sum_{j=1}^{n} p_j \log_2 p_j \qquad (15)$$

Where $p_j$ identifies frequencies related with possible sequence patterns and $n$ expresses permutation order ($n>=2$).

Permutation entropy is a measure for nonstationary, nonlinear and chaotic time series in the presence of dynamical noise. It has robust and fast results disregarding of noise. This measure has low computational complexity and therefore is suitable for the analysis of large datasets [45].

In this work, we selected the embedding dimension of 3 and time delay of 1.

The wavelet packet is applied to calculate the following five entropies. These entropies are defined as [46]:

Norm entropy (NormEn)

$$\text{NormEn}=\sum_i |x_i|^p \text{ , P}\geq 1 \qquad (16)$$

Threshold entropy (ThreshEn)

$$\text{ThreshEn} = \{\text{i such that } |x_i|>P\} \text{ (P>0)} \qquad (17)$$

This expresses the time instants number when the signal is larger than the threshold.

The threshold value is selected at 0.2.

Sure entropy (SureEn)

$$E_{SEn} = N - \#\{i \textit{ such that } |x_i| \le \varepsilon\} + \sum_i \min(x_i^2, \varepsilon^2) \qquad (18)$$

Log Energy entropy (LogEn)

$$E_{LgEn} = \sum_{i=1}^{N} \log(x_i^2) \qquad (19)$$

Where $x$ is the signal and $x_i$ denote coefficients of $x$ in the orthonormal basis. Also $p$, $P$ and $N$ represent power, threshold value and signal length, respectively.

**Classification**

Support Vector Machine (SVM)

SVM is a pattern recognition algorithm improving ideally the generalization capability of a learning machine due to minimizing the principle of structural risk. The SVM method maps the data to an n-dimensional space where $n$ is the number of features. It then uses a decision boundary to separate the two classes. To achieve this, the margin between the hyperplanes should be maximized.

In this study, SVM and LDA classifiers have been used. SVM identifies complicated patterns in data and in spite of having many dimensions it avoids over-fitting [47].

Linear Discriminant Analysis (LDA)

LDA classifier creates a new variable that is a combination of the main predictors, which is achieved by maximizing the difference between the predefined groups with respect to the new variable. According to this algorithm, predictors scores are combined to form a new variable with a different score. At the end of the process, each class has a normal distribution of distinct scores with the largest possible difference between the mean of classes score.

The basic assumption of the LDA is that the independent variables have a normal distribution and it operates based on the linear combination of the variables. This classifier calculates the average vector of each class. Then, when an observation point is close to average vector of a class in the discriminant variables space, the above-stated observation will be assigned to that class [48].

**Our proposed method**

Figure. 1 indicates a flowchart of our analysis.

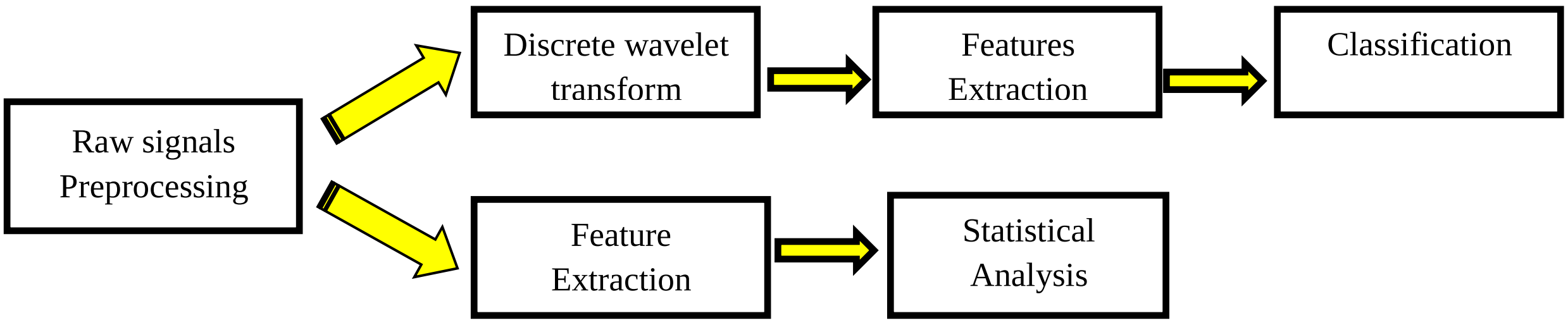

Figure. 1. The block diagram of the proposed method for discriminating epileptic and normal EEG data.

In this paper, two datasets of the healthy group with open eyes and epileptic patients group were filtered using the Butterworth filter with a cutoff frequency of 0.5 to 40 Hz. The signal was then decomposed into frequency sub-bands using the Discrete Wavelet Transform (DWT). DWT is a linear time-frequency method that allows local signal analysis with high resolution. Therefore, this is an appropriate method for nonstationary signals.

DWT decomposes the signal to *n* levels, where the approximation and detail coefficients of each level can be calculated. The signal was decomposed into 5 levels. For this purpose, the Daubechies4 wavelet filter was used.

The following features were then extracted from the signal wavelet coefficients:

The embedding dimension and embedding delay of the state space and the 10 types of entropies including AppEn, SampEn, ShanEn, FuzzyEn, SEn, PermEn, ThreshEn, NormEn, SureEn, LogEn were extracted from each wavelet subband of EEG signals.

Classification

After extracting the features from all sub-bands, the signal classification was performed using LDA and SVM classifiers.

K-fold cross validation (K=10) was applied to ensure the performance of the classifier and to prevent over-fitting. The performance of the classifiers has been evaluated by testing classification accuracy, specificity and sensitivity.

Statistical analysis

In this research, we analyzed all the features extracted from EEG signals with the statistical test to evaluate the difference of significance level between the two groups. Features were examined using the Mann-Whitney U test by determining the p-value between epileptic patients and the normal group. Also, the mean and standard deviation of extracted features were calculated for each group and the results were compared.

**Results**

Statistical test results for embedding parameters and 10 sets of predefined entropies between healthy and patient groups are shown in Tables 2 & 3, respectively.

Mean and standard deviation for all of the features are calculated for both classes. The normal group had lower embedding parameters values than the epileptic group but higher entropy values as listed in Tables 2 & 3.

Performance measures of LDA and SVM classifiers have been shown in tables 4 to 7. These results obtained by investigating each feature separately.

According to classifiers performance for diagnosing epileptic patients, SVM achieved the highest accuracy, in the classification of EEG signals (see Tables 6 & 7).

Table 2. Mean and Standard Deviation (SD) of embedding parameters extracted from healthy and patient groups.

| **Features** | **Healthy** | | **Patient** | | **p-Value** |
|---|---|---|---|---|---|
| | **Mean** | **SD** | **Mean** | **SD** | |
| **Embedding delay** | 6.900 | 1.040 | 12.510 | 6.752 | 4.48e-24 |
| **Embedding dimension** | 7.140 | 0.532 | 8.020 | 1.054 | 2.21e-04 |

Table 3. Mean and Standard Deviation (SD) of entropy measures extracted from healthy and patient groups.

| **Features** | **Healthy** | | **Patient** | | **p-Value** |
|---|---|---|---|---|---|
| | **Mean** | **SD** | **Mean** | **SD** | |
| **Approximate Entropy** | 0.891 | 0.106 | 0.900 | 0.117 | 1.02e-01 |
| **Sample Entropy** | 0.060 | 0.010 | 0.005 | 0.000 | 2.8e-02 |
| **Permutation Entropy** | 1.369 | 0.040 | 1.049 | 0.093 | 0.004 |
| **Fuzzy Entropy** | 0.030 | 0.012 | 0.004 | 0.002 | 2.56e-03 |
| **Shannon Entropy** | -59e+07 | 2e+07 | -60e+7 | 5e+09 | 5.76e-02 |
| **Spectral Entropy** | 0.777 | 0.009 | 0.756 | 0.015 | 4.8e-02 |
| **NormEn** | 7.7 e+03 | 2e+01 | 6.17 e+02 | 1e+01 | 1.81e-05 |
| **ThreshEn** | 5.7 e+05 | 3.4e+03 | 1.7 e+03 | 2.6e+01 | 9.01e-06 |
| **LogEn** | 9.3 e+05 | 1.6e+03 | 6.7 e+03 | 1.9e+01 | 3.27e-06 |
| **SureEn** | 8.4 e+02 | 2.4e+01 | 9.37 e+01 | 2.1e+01 | 2.76e-05 |

Table 4. The performance of LDA classifier for embedding parameters extracted from healthy and patient groups.

| **Performance criteria** | **Embedding dimension** | **Embedding Delay** |
|---|---|---|
| Accuracy | 95.85 | 97.78 |
| Specificity | 96.91 | 95.31 |
| Sensitivity | 94.79 | 100 |

Table 5. The performance of LDA classifier for entropy measures extracted from healthy and patient groups.

| Performance criteria | AppEn | SampEn | PermEn | FuzzyEn | ShanEn | SpectralEn | NormEn | ThreshEn | LogEn | SureEn |
|---|---|---|---|---|---|---|---|---|---|---|
| Accuracy | 86.11 | 100 | 81.67 | 90.56 | 77.78 | 83.33 | 98.89 | 100 | 100 | 98.89 |
| Specificity | 82.09 | 100 | 80.93 | 92.01 | 74.44 | 86.30 | 98.06 | 100 | 100 | 97.78 |
| Sensitivity | 90.35 | 100 | 82.93 | 89.74 | 82.51 | 90.72 | 100 | 100 | 100 | 100 |

Table 6. The performance of SVM classifier for embedding parameters extracted from healthy and patient groups.

| Performance criteria | Embedding dimension | Embedding Delay |
|---|---|---|
| Accuracy | 96.67 | 99.44 |
| Specificity | 95.54 | 99.07 |
| Sensitivity | 97.88 | 100 |

Table 7. The performance of SVM classifier for entropy measures extracted from healthy and patient groups.

| Performance criteria | AppEn | SampEn | PermEn | FuzzyEn | ShanEn | SpectralEn | NormEn | ThreshEn | LogEn | SureEn |
|---|---|---|---|---|---|---|---|---|---|---|
| Accuracy | 90.56 | 100 | 97.78 | 99.44 | 85.56 | 86.11 | 100 | 100 | 100 | 100 |
| Specificity | 91.06 | 100 | 96.13 | 98.89 | 80.25 | 82.09 | 100 | 100 | 100 | 100 |
| Sensitivity | 90.49 | 100 | 100 | 100 | 93.03 | 90.35 | 100 | 100 | 100 | 100 |

## Discussion

First, we extracted embedding parameters and 10 sets of entropies from each wavelet sub-band. Next, we used statistical analysis that revealed which features were statistically significant. We also fed them to SVM and LDA classifiers. Our results showed measures of Embedding Dimension, Embedding Delay, PermEn, FuzzyEn, SampEn, NormEn, SureEn, LogEn and ThreshEn can accurately discriminate epileptic patients from normal subjects. These features were statistically significant between the two groups. The p-value of them revealed their ability for significant differentiation (p-value<0.05).

The probable reasons of the high correct classification rate of EEG signals with the mentioned features are as follows:

Embedding parameters are important factors for phase space reconstruction. Embedding dimension shows the lowest number of uncorrelated orientations in the phase space. Over-embedding increases redundancy of time series and maximizes the information rate of reconstructed phase space. Two methods of AMI and FNN are used to calculate the embedding delay and embedding dimension, respectively. AMI approach in comparison with the other methods such as method based on singular value decomposition (SVD) [49] considers the nonlinear interrelation which ensures the independence criterion that guarantees the uncorrelated property of orientations and leads to the better reconstruction of the state space. But the method based on SVD minimizes only the linear correlation between components which doesn’t mean independence.

According to Eq. (1) more delayed time series are involved in the state-space vector generated for epileptic EEG signals. So it is expected that the calculated parameters will be higher than those of normal EEG signals. This is because the normal EEG signals are inherently random but change to chaos deterministic dynamic state during epileptic seizures [50]. The results also indicate the difference between the dynamics of EEG signals, and the values of both state-space features obtained for the patient's EEG signals were higher than those of the healthy group (see Table 2).

Entropy is an appropriate tool to measure the complexity of EEG signals. Epileptic EEG signals are usually characterized by rhythmic patterns. Rhythmic activity during epileptic seizures leads to less complexity in epileptic EEG signal in contrast with normal EEG signal [50]. The decrease of complexity can be quantified with a nonlinear measure like entropy. It is expected that the entropy decreases in epileptic patients compared with the healthy group. The results shown in Table 3 confirm the validity of this statement.

AppEn, SampEn, fuzzyEn are appropriate measures for separating different systems such as periodic, multi periodic, chaotic and stochastic when the time series is noisy. They have wide range of applications for the characterization of nonlinear signals. A periodic time series has relatively low entropy since it comprises of many repetitive patterns. However, in stochastic time series, a pattern observed in one observation vector is not likely repeated throughout others and the entropy is higher [50]. Since the epileptic EEG signals are more periodic than the normal ones, the values calculated for the mentioned entropy measures of the patient group were expected to be lower than the normal group as the results are also consistent with this expectation.

As mentioned before, AppEn is dependent on the length of the signal and suffers from bias as well as considering self-matches [42], but in SampEn and FuzzyEn, there are no such issues[51]. Additionally, AppEn has a rigid boundary of similarity which is highly sensitive to slight changes in distance between two vectors [52]. In other hand, FuzzyEn as a more precise feature has a soft boundary and evaluates the similarity degree of two vectors fuzzily [44]. Above statements are consistent with the obtained results and suggests SampEn and FuzzyEn as better measures than AppEn.

PermEn is a robust and simple measure of nonstationary, nonlinear and chaotic time series disregarding of noise. It can be used for the processing of big datasets without the need for fine-tuning of the parameters [45]. Also, SpectralEn, NormEn, SureEn, ThreshEn, and LogEn have the ability of fine variability detection for nonstationary signals and require little computation time [53]. Then, due to the descriptions and obtained results, the mentioned entropies can be considered as effective measures for the extraction of signal information. On the other hand, ShanEn has few basic disadvantages: probability of over-estimation of the entropy level due to a large number of areas used and also inability in illustration of the temporal relation between the various values of the time series [54].

Entropy increase of normal EEG signal can be attributed to the stochastic dynamics of the signal which increases irregularity and randomness of time series. Epileptic EEG signal is more regular and predictable than normal EEG signal. Hence, epileptic patients are expected to have lower EEG entropy consistent with our obtained results [53].

It is noteworthy that in this study all the extracted features, have been studied separately. In general, the size of the feature vector is a challenging problem for the classification algorithms. The high dimension of feature space increases the complexity of the model and results in more time consumption of both training and test along with the increase of redundancy of features which leads to lower accuracy in estimation of the model parameters. If a single feature can guarantee a reliable classification performance, there will be no need for using multiple features.

LDA classifier separated two classes with high accuracy for measures of Embedding delay, Embedding Dimension, SampEn, NormEn, SureEn, LogEn and ThreshEn. Therefore, we were able to find the features that can classify datasets even with a simple and linear classifier like LDA.

Although LDA had poor accuracy to classify datasets for features PermEn, FuzzyEn, SVM performed better and had high accuracy (see Tables 4 & 5).

In the current study, all of the measures except ShanEn, SpectralEn and AppEn showed the appropriate performance as a single feature for discriminating epileptic subjects from normal individuals. In fact, the classifiers achieved reliable accuracy using these features. Our results are in agreement with previous studies that show the significant alteration of these features in epileptic patients. Also, SVM based on measures of SampEn, NormEn, SureEn, LogEn and ThreshEn achieved classification accuracy of 100%.

## Conclusion

SVM and LDA classifiers have been applied to classify EEG signals in two groups of healthy and epileptic patients. By comparing the sensitivity, specificity, and accuracy of these classifiers, they yielded reliable results by effectively discriminating the EEG signals of epileptic patients from those of normal subjects. Due to nonlinearity and nonstationary properties of the EEG signal, nonlinear analysis can be useful for reliable quantifying of classification and identifying abnormal activities of the EEG signals.

## Future work

The examined features can be employed in other biomedical applications such as detection of eye diseases, cardiac abnormalities, alzheimer's disease, and autism. Also in this work, the dataset classified between only two groups. It is believed that these features could be beneficial for the categorization of more groups in the future. Also, these features can be applied to the other Epileptic EEG datasets to evaluate the generalization of their performance.

## Acknowledgments

We would like to appreciate Anderzejak et al. from the University of Bonn, Germany for making the raw EEG data accessible. Finally, we thank Ali Pouresmaeil for his comments which greatly improved the manuscript.